# Comment on *μp* and *μA* Luminosities of Muon-Ion Collider at BNL

Burak Dagli[1], Bora Ketenoglu[2,*] and Saleh Sultansoy[1,3]

[1] Department of Materials Science and Nanotechnology Engineering, TOBB University of Economics and Technology, Ankara, Turkey
[2] Department of Engineering Physics, Ankara University, Ankara, Turkey
[3] ANAS Institute of Physics, Baku, Azerbaijan

[*]Corresponding Author: bketen@eng.ankara.edu.tr

**Abstract**

Luminosities of muon-proton and muon-nucleus collisions at a recently proposed muon-ion collider (MuIC) at Brookhaven National Laboratory (BNL) in the USA have been estimated using "A Luminosity Optimizer for High Energy Physics (AloHEP)" software keeping in mind beam-beam tune-shift values. It is shown that $L_{\mu p} \approx 3.6 \times 10^{31}$ cm$^{-2}$s$^{-1}$ and $L_{\mu\text{-Au}} = 2.84 \times 10^{27}$ cm$^{-2}$s$^{-1}$ values can be achieved at MuIC. The physics search potential of the collider is superior to HERA. However, LHeC will provide 2 orders of higher luminosities at approximately same center-of-mass (CoM) energies.

Keywords: Muon-hadron colliders, MuIC, RHIC, Luminosity, Beam-beam tune-shift

## 1. Introduction

Recently, a muon-ion collider at BNL has been proposed [1] as a path to energy frontier $\mu^+\mu^-$ collider (for earlier *μp* and *μA* collider proposals see [2,3] and references therein). Later, MuIC at BNL has been considered in [4] and [5] as well. Even though Reference [1] provides perfect information about physics search potential of the MuIC collider, luminosity of *μp* collisions are over-estimated and luminosity estimations for *μA* collisions are not provided. Furthermore, beam-beam tune-shift effects have been ignored. In this comment, we consider luminosities of *μp* and *μA* options of the MuIC at BNL in detail.

## 2. Luminosity of *μp* collisions

Main parameters of the MuIC at BNL are given in Table 1 (Table 2 of Reference [1]). Here, we have implemented an additional row for beam-beam tune-shifts.

**Table 1.** Proposed parameters of MuIC at BNL *μp* option.

| Parameters | Muon | Proton |
|---|---|---|
| Beam Energy [TeV] | 0.96 | 0.275 |
| CoM Energy [TeV] | 1.03 | |
| Bunch intensity [$10^{11}$] | 20 | 3 |
| Norm. emittance, $\varepsilon_{x,y}$ [μm] | 25 | 0.2 |
| $\beta^*$ at IP [cm] | 1 | 5 |
| Trans. beam size, $\sigma_{x,y}$ [μm] | 5.2 | 5.8 |
| Muon repetition rate, $f_{rep}$ (Hz) | 15 | |
| Cycles/collisions per muon bunch | 3279 | |
| $L_{\mu p}$ [$10^{33}$cm$^{-2}$s$^{-1}$] | 7 | |
| Beam-beam tune-shifts, $\xi$ | 0.013 | 1.2 |

Implementing parameters from Table 1 into AloHEP, we obtain L = $6.9 \times 10^{33}$cm$^{-2}$s$^{-1}$ without muon decays, which is well-consistent with the luminosity in Table 1, and L = $4.35 \times 10^{33}$cm$^{-2}$s$^{-1}$ with muon decays. AloHEP results for beam-beam tune-shift values are $\xi_p = 1.2$ and $\xi_\mu = 0.013$. It is seen that beam-beam tune-shift value for proton beam is unacceptably high while it is reasonable for muon beam.

Beam-beam tune-shift value for the proton beam can be reduced to an acceptable value of 0.01 by decreasing number of muons per bunch by 120 times, which will result in a corresponding reduction of luminosity as a matter of course. Therefore, realistic value for $\mu p$ luminosity at MuIC is $L_{\mu p} \approx 3.6 \times 10^{31}$ cm$^{-2}$s$^{-1}$.

## 3. Luminosity of $\mu A$ collisions

For estimation of luminosity of the RHIC based muon-nucleus colliders, let us consider $Au$ option. Corresponding parameters are summarized in Table 2 (see Table 32.5 in [6]).

**Table 2.** Parameters of Gold beam

| Beam Energy [TeV] | 19.7 |
|---|---|
| Circumference [km] | 3.834 |
| Particle per Bunch [$10^9$] | 2 |
| Revolution rate [Hz] | 78250 |
| Normalized Emittance [µm] | 2.23 |
| β* function at IP [cm] | 70 |
| Bunches per beam | 111 |

Implementing muon beam parameters from Table 1 and $Au$ beam parameters from Table 2 into AloHEP software, we have obtained parameters of $\mu$-$Au$ collisions given in Table 3.

**Table 3.** RHIC based $\mu$-$Au$ collider parameters

| √s [TeV] | 8.7 |
|---|---|
| IP beam size [µm] | 115 |
| Tune shift $\xi_{Au}$ | 3.82 |
| Tune shift $\xi_\mu$ [$10^{-3}$] | 6.85 |
| Peak L [$10^{29}$cm$^{-2}$s$^{-1}$] | 1.18 |
| L with decay [$10^{29}$cm$^{-2}$s$^{-1}$] | 0.74 |

It is seen that beam-beam tune-shift parameter for Gold beam is very high. It can be reduced to an acceptable value of 0.01 by decreasing number of muons 382 times per bunch, which will result in a corresponding reduction in luminosity as a matter of course. This reduction may partially be compensated by increasing the number of $Au$ per bunch. Increase of this number by a factor of 14.6 results in tune-shift value of 0.1 for muon beam, which is acceptable for $\mu$-$Au$ collider. In summary, these modifications lead to 26 times reduction of luminosity values (see Table 4).

**Table 4.** RHIC based $\mu$-$Au$ collider parameters with upgraded numbers of $\mu$ and $Au$ per bunch

| √s$_{\mu\text{-}Au}$ [TeV] | 8.7 |
|---|---|
| N$_\mu$ [$10^9$] | 5.24 |
| N$_{Au}$ [$10^{10}$] | 2.92 |
| IP beam size [µm] | 115 |
| Tune shift $\xi_{Au}$ | 0.01 |
| Tune shift $\xi_\mu$ | 0.1 |
| Peak L$_{\mu\text{-}Au}$ [$10^{27}$ cm$^{-2}$s$^{-1}$] | 4.50 |
| L$_{\mu\text{-}Au}$ with decay [$10^{27}$ cm$^{-2}$s$^{-1}$] | 2.84 |

Therefore, realistic value for $\mu p$ luminosity at MuIC is $L_{\mu p} \approx 2.8 \times 10^{27}$ cm$^{-2}$s$^{-1}$.

## 4. Conclusion

As mentioned in Introduction section, Reference [1], where MuIC have been proposed, contains excellent review of physics search potential of $\mu p$ and $\mu A$ colliders. But luminosity part is incomplete: $L_{\mu p}$ is over-estimated, $L_{\mu A}$ is not provided at all.

In this study, we show that a realistic luminosity value for $\mu p$ collisions at the MuIC is $L_{\mu p} \approx 3.6 \times 10^{31}$ cm$^{-2}$s$^{-1}$, which is comparable with the HERA $ep$ collider luminosity. For this reason, main advantage of MuIC $\mu p$ option over HERA is three times higher center-of-mass energy. Because of the fact that HERA did not offer $eA$ collisions, MuIC (which contains both $\mu p$ and $\mu A$ options) will provide superior physics search potential when compared with HERA.

As known, CERN has planned to construct LHeC [7] $ep$ collider based on 60 GeV electron beam from ERL and 7 TeV proton beam from LHC. The LHeC with √s = 1.3 TeV is comparable with MuIC $\mu p$ option, while luminosity value $L_{ep} = 10^{33\text{-}34}$ cm$^{-2}$s$^{-1}$ is 2 orders higher. Therefore, since LHeC offers both $ep$ and $eA$ options, its physics search potential is much superior to that of MuIC.

Finally, let us mention that decreasing of number of muons by 2 orders (which is compulsory to keep proton beam tune-shift under control) significantly reduces both radiation hazard problem caused by neutrinos from muon decays and muon production rate problem. Therefore, MuIC can be truly considered as a "path to a new energy frontier of $\mu^+\mu^-$ colliders" as emphasized in [1].